\documentclass[aps,prb,twocolumn,showpacs,citeautoscript,superscriptaddress,longbibliography]{revtex4-1}

\usepackage{graphicx}  
\usepackage{epstopdf}
\usepackage{dcolumn}   
\usepackage{bm}        
\usepackage{amsmath}
\usepackage{amssymb}   
\usepackage{wrapfig}
\usepackage{array}
\usepackage[caption=false]{subfig}
\usepackage[bookmarks=false,linkcolor=blue,urlcolor=blue,colorlinks,citecolor=blue]{hyperref}
\usepackage{exscale,relsize}
\usepackage{color}
\usepackage{times, verbatim}
\usepackage{pstricks}
\usepackage{bbold}

\usepackage{pst-all}

\newcommand{\bra}[1]{\langle #1|}
\newcommand{\ket}[1]{|#1\rangle}

\DeclareMathOperator{\tr}{tr}

\DeclareMathOperator{\rhoMZM}{\rho_{\text{MZM}}}

\makeatletter

\begin{document}

\title{Dephasing of Majorana-based qubits}

\author{Christina Knapp}
\affiliation{Department of Physics, University of California, Santa Barbara,
	California 93106 USA}

\author{Torsten Karzig}
\affiliation{Station Q, Microsoft Corporation, Santa Barbara, California 93106-6105 USA}

\author{Roman M. Lutchyn}
\affiliation{Station Q, Microsoft Corporation, Santa Barbara, California 93106-6105 USA}

\author{Chetan Nayak}
\affiliation{Station Q, Microsoft Corporation, Santa Barbara, California 93106-6105 USA}
\affiliation{Department of Physics, University of California, Santa Barbara,
	California 93106 USA}

\begin{abstract}
We analyze charging-energy-protected Majorana-based qubits, focusing on the residual dephasing
that is present when the distance between Majorana zero modes (MZMs) is insufficient for full
topological protection. We argue that the leading source of dephasing is
 $1/f$ charge noise.
This noise affects the qubit as a result of the hybridization energy and charge distribution associated with weakly-overlapping MZMs,
which we calculate using a charge-conserving formalism.
We estimate the coherence time to be hundreds of nanoseconds for Majorana-based qubits whose MZM separation is $L\sim 5\xi$ (with $\xi$ being the coherence length). The coherence time grows exponentially with MZM separation and eventually becomes temperature-limited for $L/\xi \sim 30$.
\end{abstract}

\date{\today}

\maketitle

\section{Introduction}

Topological phases offer the promise of qubits that are insensitive to local sources of noise, provided that the relevant distance and time scales
are sufficiently large~\cite{Kitaev97,Nayak08}. When qubit operations are done too rapidly, however, diabatic errors can occur~\cite{Cheng11,Karzig13,Scheurer13,Knapp16}.  Furthermore, as the separation between topological excitations is decreased, eventually approaching and then
falling below the coherence length, topological qubits evolve smoothly
into more conventional (local) qubits and are susceptible to the same noise sources~\cite{Rainis12,Goldstein11,Budich12}. At present, the most promising approach to topological quantum computing encodes the qubit in the joint parity state of Majorana zero modes (MZMs), exotic defects of topological superconductors that obey non-Abelian statistics~\cite{Read00,Kitaev01,Alicea12a}.  A one-dimensional topological superconductor can be engineered out of a semiconductor nanowire with strong spin orbit coupling, proximitized by an $s$-wave superconductor and subjected to a magnetic field~\cite{Sau10a,Lutchyn10,Oreg10}.
Motivated by the strong experimental evidence for the observation of MZMs in such systems~\cite{Mourik12,Rokhinson12,Deng12,Churchill13,Das12,Finck12,Albrecht16, Zhang2017, Lutchyn17}, there is a growing interest in moving beyond detection of MZMs to their application in topological quantum computing~\cite{Aasen16,Lutchyn17}.  In particular, recent theoretical work has proposed
qubits comprised of four or six MZMs on an island with substantial charging
energy~\cite{Plugge16b,Karzig17}. In this paper, we analyze the dephasing of such qubits
that occurs when two of the MZMs on such an island approach each other.

The MZM qubits of Refs.~\onlinecite{Plugge16b,Karzig17} have a fixed electric charge, which protects them from poisoning by excited quasiparticles originating elsewhere in the device. However, they are still vulnerable to two types of errors.
(1) An excited fermionic quasiparticle on the island can be absorbed or emitted by a MZM. If this happens once,
it takes the qubit out of the computational subspace; if it happens twice, it causes a bit or phase error, depending on which two MZMs are affected. (2) When the separation between MZMs is not large compared to the coherence length, the overlap between
MZMs causes a redistribution of the electric charge in the island. The resulting charge distribution
(which will, in general, have a non-vanishing line dipole moment between the semiconductor and superconductor) couples to phonons and
the electrostatic environment of the island. These low-energy degrees of freedom cause
the qubit to decohere.

In this paper, we give quantitative estimates
for both types of errors mentioned in the previous
paragraph. Type (1) depends on the density of excitations, and therefore is small when this density is
small.  In thermal equilibrium, these errors are exponentially suppressed in the product of the gap $\Delta$ and the inverse temperature $\beta$. The main focus of the paper is to quantify type (2) errors by computing the hybridization energy and charge distribution associated with MZMs using a charge-conserving formalism.
We show how a dipole moment develops between a semiconductor nanowire and its superconducting
shell. This dipole formation is analogous to the situation that occurs in a double quantum dot charge qubit, except that
the transferred charge is much less than the charge of an electron. We give quantitative estimates of the resulting dephasing using measurements of the electrostatic noise spectrum in similar devices and the electron-phonon coupling and phonon spectrum of InAs.
Very similar physics applies to the measurement process proposed in Ref.~\onlinecite{Karzig17}:
when a quantum dot is coupled to a MZM, a dipole moment develops between the quantum
dot and the qubit. We report the corresponding dephasing times which quantify how fast the environment reads out the parity of a pair of MZMs during the measurement process.

The remainder of this paper is organized as follows.  In Section~\ref{sec:setup}, we develop the basic setup of the qubit-environment coupling.  In Section~\ref{sec:charge-distribution}, we calculate the hybridization energy and charge distribution associated with the overlap of a pair of MZMs.  We estimate qubit dephasing times due to several different noise sources in Section~\ref{sec:dephasing-times}. In Section~\ref{sec:discussion}, we discuss additional effects of charge noise on the qubit system. We conclude in Section~\ref{sec:conclusions}. Details of the various discussions are relegated to the appendices.

\section{Basic Setup}\label{sec:setup}

Consider a two-level system with density matrix $\rho(t)$, described by a Hamiltonian $H_S=\Omega \sigma^z$.  We assume that the system interacts weakly with its environment, described by a Hamiltonian $H_E$, and that the environment is in thermal equilibrium at inverse temperature $\beta$.  The density matrix $\rho(t)$ undergoes a particularly simple time evolution when the system-environment interaction is diagonal in the system's energy basis,
\begin{equation}
\label{eqn:H_SE}
H_\text{SE} = \frac{a_z}{2}{\sigma^z}\otimes \Phi,
\end{equation}
where $\Phi$ acts on the environment degrees of freedom.  The diagonal elements of $\rho(t)$ have constant magnitude and the off-diagonal elements decay according to~\cite{Cywinski2008}
\begin{equation}
\label{eq:dephasing}
|\rho_{01}(t)| = e^{-{B^2}(t)} |\rho_{01}(0)|,
\end{equation}
where
\begin{equation}
\label{eq:basic-dephasing}
{B^2}(t) \equiv a_z^2 \int_{0}^{\infty} d\omega \, S(\omega)\, \frac{{\sin^2}(\omega t/2)}{(\omega/2)^2}.
\end{equation}
Here, the noise spectral function of $\Phi$ is given by
\begin{equation}\label{eq:S-omega}
S(\omega) \equiv \int_{-\infty}^{\infty} \, dt\, \frac{e^{i\omega t}}{2\pi}\left( \frac{
\langle \Phi(t) \Phi(0) \rangle+ \langle \Phi(0)\Phi(t)\rangle}{2}\right)
\end{equation}
where  ${\langle \Phi(t) \Phi(0) \rangle \equiv
\text{tr}\left(e^{-\beta {H_\text{E}}} \Phi(t) \Phi(0) \right)}$.

\begin{figure}
\includegraphics[width=\columnwidth]{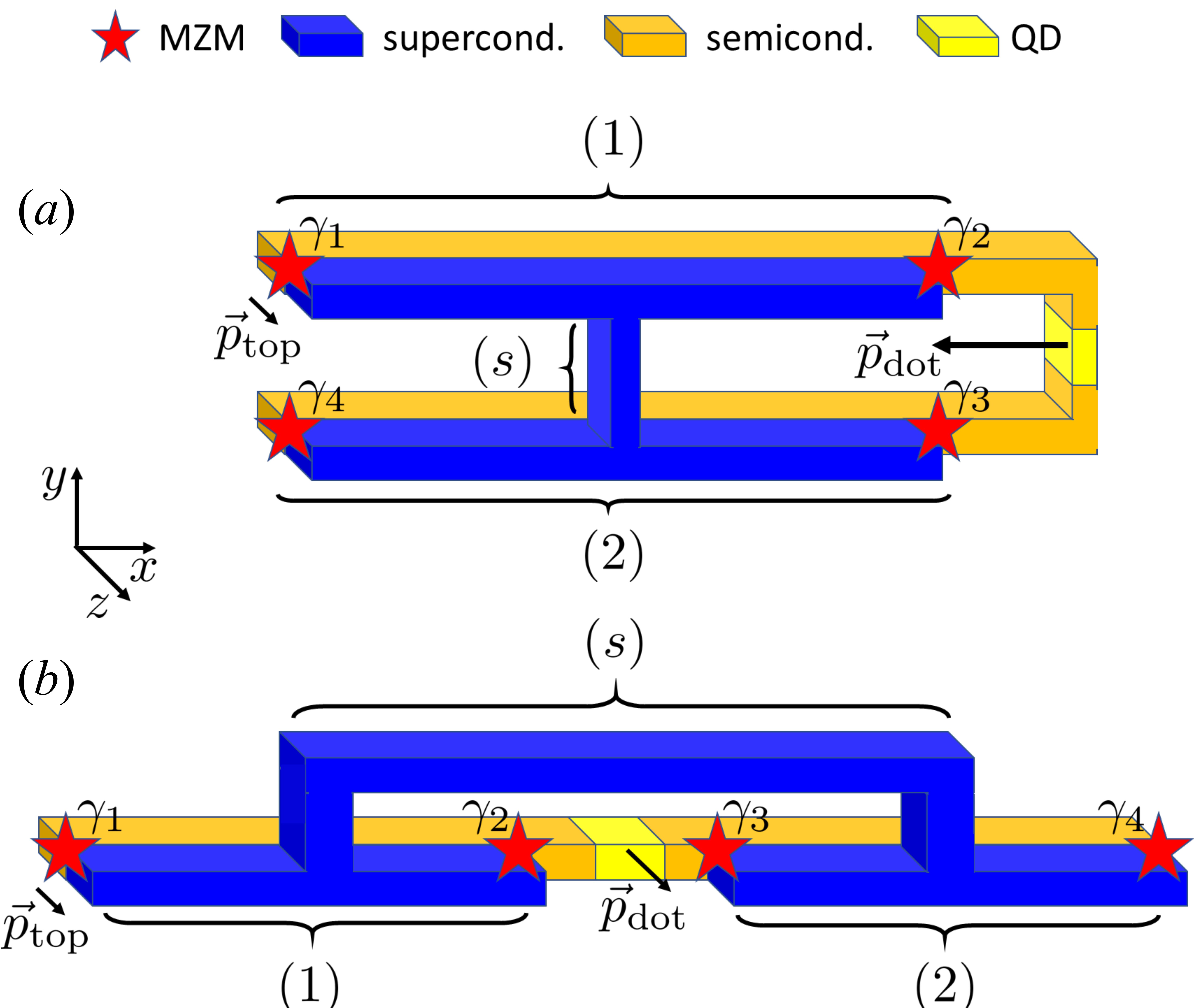}
\caption{{\it (Color online)} Two charge-protected MZM qubit geometries: $(a)$ Two-sided tetron and $(b)$ loop qubit.   Each design has two topological sections, labeled $(1)$ and $(2)$, consisting of a semiconducting nanowire (orange) proximitized by a superconducting wire (blue), and tuned into the topological phase so that MZMs (red stars) are localized at either end.  The two topological sections are connected by a trivial superconductor, the ``superconducting backbone,'' labeled by $(s)$.  The superconducting backbone ensures that the device acts as a single superconducting island, thereby allowing superpositions of all (total fermion parity even) MZM states. When the superconducting island has appreciable charging energy, extrinsic quasiparticle poisoning is strongly suppressed, hence the designation that these are ``charging-energy protected MZM qubits.''  MZMs belonging to the same wire ($\gamma_1$ and $\gamma_2$ or $\gamma_3$ and $\gamma_4$) will slightly overlap, resulting in a relative charge distribution between the semiconductor and superconductor in the topological sections.  This charge buildup results in a dipole moment, $\vec{p}_\text{top}$, oriented perpendicular to the semiconductor/superconductor interface.  Provided the lengths of the topological wires are equivalent in the two designs, $\vec{p}_\text{top}$ will be the same.  A measurement of the fermion parity $i\gamma_2\gamma_3$ is performed by tunnel coupling MZMs 2 and 3 to an auxiliary quantum dot (yellow), located in the semiconducting region connecting wires $(1)$ and $(2)$.  The qubit-quantum dot system also forms a dipole moment, $\vec{p}_\text{dot}$, whose magnitude and direction depends on the device geometry.   We assume an essentially vanishing screening length so that the displacement vector entering $\vec{p}_\text{dot}$ points from the quantum dot to the surface of the superconductor: note that this results in $\vec{p}_\text{top}$ and $\vec{p}_\text{dot}$ being parallel in $(b)$, provided topological sections $(1)$ and $(2)$ are equidistant from the quantum dot.  Coupling of $\vec{p}_\text{top}$ to the environment sets the dephasing time of the qubit; coupling of $\vec{p}_{\text{dot}}$ to the environment sets how fast the environment measures $i\gamma_2\gamma_3$.      
}
\label{fig:qubit}
\end{figure}

We use Eq.~(\ref{eq:basic-dephasing}) to analyze dephasing times in charge-protected MZM qubits.  In Fig.~\ref{fig:qubit}, we depict two possible geometries of such qubits. The common elements of both geometries are: two topological sections built from a semiconductor wire (light orange) proximitized by a superconductor (dark blue), connected by a trivial $s$-wave superconductor [labeled $(s)$] to form a Coulomb-blockaded superconducting island hosting four MZMs. We call the trivial superconducting region the ``backbone.''  The qubit is encoded according to $\sigma^z \equiv i \gamma_1 \gamma_2$ (note that in the ground state, $i\gamma_1\gamma_2=i\gamma_3\gamma_4$). The main difference between the two geometries is that the upper design $(a)$ requires at least two semiconducting nanowires while the lower design $(b)$ can be realized with a single nanowire and a loop-shaped backbone.

We consider the limit in which the energy gap in the superconducting backbone is much larger than in the topological sections.  Then, the amplitude for a fermion to tunnel from $\gamma_1$ or $\gamma_2$ to $\gamma_{3}$ or $\gamma_4$ will be very small.  The dominant error mechanism will be dephasing from the coupling of the electromagnetic environment to the charge shared by $\gamma_1$ and $\gamma_2$ (and shared by $\gamma_3$ and $\gamma_4$). This assumption simplifies our calculations, but does not change our main results.

The qubit states stored in $i\gamma_1\gamma_2$ are slightly split in energy by $\varepsilon_{\text{hyb}}$, resulting from overlap of the MZM wavefunctions.  This hybridization energy fluctuates with the electromagnetic environment, resulting in the dephasing of the qubit.  The qubit-environment coupling can be modeled by the simple Taylor expansion:
\begin{equation}\label{eq:HMZME}
\begin{split}
H_\text{MZM-E} &=\frac{1}{2}  \left( \frac{\partial \varepsilon_{\text{hyb}}}{\partial E_z}\right) i\gamma_1 \gamma_2 \otimes \delta E_z(t)
\end{split}
\end{equation}
where $E_z$ is the electric field component perpendicular to the semiconductor-superconductor interface, as shown in Fig.~\ref{fig:qubit}.  This interaction can equivalently be understood as the electrostatic environment coupling to the dipole moment
\begin{equation}\label{eq:p-top}
\vec{p}_\text{top}= \frac{\partial \varepsilon_\text{hyb}}{\partial E_z} \hat{z},
\end{equation}
whose sign depends on the parity of MZMs $\gamma_1$ and $\gamma_2$.  We calculate the hybridization energy and the charge distribution in the topological wire leading to this dipole moment in Section~\ref{sec:charge-distribution}.

For both qubit designs shown in Fig.~\ref{fig:qubit}, a measurement is performed by coupling two of the MZMs to an auxiliary quantum dot (yellow)~\cite{Karzig17}.  This coupling can be achieved by lowering tunnel barriers (not shown) in the semiconducting region neighboring MZMs $\gamma_2$ and $\gamma_3$, so that an electron can tunnel into MZM $\gamma_j$ with amplitude $t_j$.  We always work in the weak coupling limit, $|t_j|\ll E_C$, where $E_C$ is the charging energy of the MZM island.  When the combined MZM qubit-quantum dot system is in its ground state, the charge distribution on the quantum dot becomes parity-dependent~\cite{Karzig17}, see Appendix~\ref{app:QD} for further details.
Measuring the quantum dot charge thus allows one to infer the MZM parity.

When the system is tuned into a measurement configuration with a single electron able to tunnel between the quantum dot and MZM qubit, another dipole moment emerges.  In the weak coupling limit, when the quantum dot and MZM qubit are off-resonant, the dipole moment is (up to corrections of order $|t_j|^2/E_C^2$)
\begin{equation}\label{eq:p-dot}
\vec{p}_\text{dot}= e\vec{d}\tau^z,
\end{equation}
where $\vec{d}$ is a displacement vector from the quantum dot to the surface of the superconductor (we assume an essentially vanishing screening length) and $\tau^z=+1$ if the electron is on the quantum dot and $-1$ if the electron is on the qubit. The qubit-dot dipole moment will couple to electromagnetic noise via
\begin{equation}
H_\text{QD-E}=\frac{1}{2}\vec{p}_\text{dot}\otimes \delta \vec{E}(t).
\end{equation}
Unlike the case of the qubit, which we want to be able to stay in a superposition for extended times, a successful measurement relies on collapsing the quantum mechanical state of the MZM island-quantum dot system. The corresponding dephasing time therefore quantifies how fast the environment measures the MZM parity $p_{23}$. Moreover, if the combined MZM island-quantum dot system populates the charge excited state during the initialization process, a short relaxation time can help to quickly return the system to its ground state.

Noise in the electromagnetic environment is given by
\begin{equation}\label{eq:SEE}
S_{E}(\omega) = \int dt  \frac{e^{i\omega t}}{2\pi}\left(\frac{ \langle \delta E_z(t)\delta E_z(0)\rangle +\langle \delta E_z(0)\delta E_z(t)\rangle}{2}\right) ,
\end{equation}
and is generally believed to be due to slow fluctuations of two level states in the environment~\cite{Paladino13, Cywinski2008}.
We do not have a microscopic model of these processes, so we extract the low-frequency form of these
fluctuations, which are assumed to have a $1/f$ frequency dependence, from experiments on
similar devices~\cite{Nakamura02,Petta04,Petersson10,Dovzhenko11,Shi13}:
\begin{equation}\label{eq:1f-spec}
S_{E}(\omega)=\frac{\alpha_E}{\omega}.
\end{equation}

Other noise sources affecting the MZM qubit are coupling to phonons and finite temperature excitations of quasiparticles in the superconductor.  The former couples to the charge distribution in the MZM qubit in much the same way as 1/$f$ charge noise, but is predicted to have a smaller effect that becomes negligible when the wires are sufficiently long, see Section~\ref{sec:dephasing-times} and Appendix~\ref{app:phonons}.  Conversely, thermally-excited quasiparticles only become a relevant noise source compared to 1/$f$ charge noise when the wire is sufficiently long such that $e^{L/\xi} \gtrsim e^{\beta \Delta}$, see Section~\ref{sec:dephasing-times} and Appendix~\ref{app:temperature}.
\\

\begin{figure}[t]
	\includegraphics[width=0.91\columnwidth]{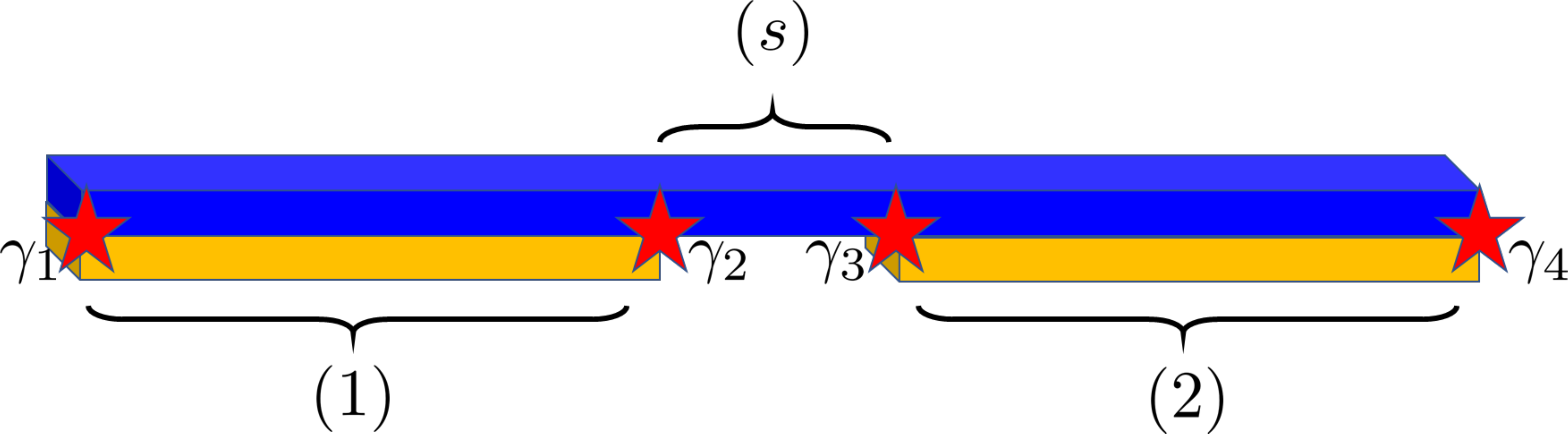}
	\caption{{\it (Color online)} Relevant geometry for the charge distribution calculation in Section~\ref{sec:charge-distribution} with the same legend as in Fig.~\ref{fig:qubit}.  Analogously to the qubit designs, there are two topological segments (each hosting a  MZM at either end) of length $L$, and a trivial superconducting region labeled $(s)$ of length $\ell$.   }
	\label{fig:geometry}
\end{figure}

\section{Hybridization Energy and Charge Distribution in MZM Qubits}
\label{sec:charge-distribution}

In this section, we calculate the hybridization energy and the charge distribution resulting from the overlap between the MZMs $\gamma_1$ and $\gamma_2$ (or equivalently between $\gamma_3$ and $\gamma_4$) in Fig.~\ref{fig:geometry}.  We expect the essential physics of this simplified geometry to be the same as that of the MZM qubits shown in Fig.~\ref{fig:qubit} when the qubit is idle (i.e., the auxiliary quantum dot is disconnected from the superconducting islands).   In order to avoid subtleties in the interpretation of the charge distribution calculated with BCS mean-field theory, we will use the explicitly charge-conserving formalism of Refs.~\onlinecite{Sau11b,Fidkowski11b,Cheng11b}.  We compare our results with previous studies of the hybridization energy~\cite{Cheng09,Cheng10} and charge distribution~\cite{Ben-Shach15,Lin-C12,Dominguez17} at the end of each subsection.

We model the topological segment $(j)$ of the device shown in Fig.~\ref{fig:geometry} as a one-dimensional spinless semiconducting wire in contact with a quasi-one-dimensional algebraically-ordered superconductor.  This model allows us to set up a controlled theory to study how phase fluctuations couple to MZMs, and ultimately to extract how the energy splitting and charge distribution depend on the fermion parity.  Electron operators in the semiconductor can be bosonized as
\begin{equation}\label{eq:psi-j}
\psi_{r}^{(j)}(x) \sim e^{-i\left(r\phi_j(x)-\theta_j(x)\right)},
\end{equation}
where $r=\pm 1$ for right or left-movers.  The superconductor electron operators are described in terms of spin ($\sigma$) and charge ($\rho$) modes
\begin{equation}\label{eq:psi}
\psi_{r,\sigma}(x)\sim e^{-\frac{i}{\sqrt{2}} \left( r\phi_\rho(x) -\theta_\rho(x) + \sigma\left(r\phi_\sigma(x)-\theta_\sigma(x)\right) \right)},
\end{equation}
where $\sigma=\pm 1$ for up or down spins and $r= \pm$ corresponds to right and left movers.  The fields $\phi_\alpha(x),~\theta_\beta(x')$ satisfy the usual commutation relations
\begin{equation}
[\partial_x \phi_\alpha(x),\theta_\beta(x')]=i\pi \delta(x-x')\delta_{\alpha \beta},
\end{equation}
for $\alpha,\beta \in \{1,2,\rho,\sigma\}$.

The above definitions yield the bosonized effective Lagrangian introduced in Ref.~\onlinecite{Fidkowski11b},
\begin{equation}
\mathcal{L}=\mathcal{L}^{(1)}+\mathcal{L}^{(2)}+\mathcal{L}^{(s)},
\end{equation}
where the trivial superconducting backbone is described by
\begin{equation}
\begin{split}
\mathcal{L}^{(s)}=\frac{1}{2\pi}\int_L^{L+\ell} dx \Big\{&-2i\left( \partial_\tau \theta_\rho \right) \left(\partial_x \phi_\rho\right) +  K_\rho v_\rho\left(\partial_x \theta_\rho\right)^2
\\ & + \frac{v_\rho}{K_\rho}\left(\partial_x\phi_\rho-k_{F}^{(\rho)}\right)^2
\Big\},
\end{split}
\end{equation}
and the topological sections are described by
\begin{widetext}
\begin{align}\label{eq:top-L}
\mathcal{L}^{(1)} &=\frac{1}{2\pi} \int_0^L dx \Big\{  -2i\left(\partial_\tau \theta_1\right) \left(\partial_x \phi_1\right) + Kv \left(\partial_x\theta_1\right)^2 +\frac{v}{ K} \left(\partial_x \phi_1 - k_{F}\right)^2
\\ &\quad \quad \quad\quad \quad ~~ -2i \left(\partial_\tau \theta_\rho\right)\left(\partial_x \phi_\rho \right) + K_\rho v_\rho\left(\partial_x \theta_\rho \right)^2 + \frac{v_\rho}{ K_\rho} \left(\partial_x \phi_\rho -k_{F}^{(\rho)} \right)^2
- \frac{\Delta_P}{ a}\cos\left(\sqrt{2}\theta_\rho - 2\theta_1\right)\Big\} \notag
\\ \mathcal{L}^{(2)} &= \int_{L+\ell}^{2L+\ell} dx \Big\{ (1) \leftrightarrow (2) \Big\}.
\end{align}
\end{widetext}
The Luttinger parameter and Fermi velocity are $K$ and $v$, respectively, for the semiconducting wires, and $K_\rho$ and $v_\rho$ for the superconductor's charge mode.
The pairing term, $\frac{\Delta_P}{2\pi a} \cos(\sqrt{2}\theta_\rho-2\theta_j)$, emerges from integrating out the gapped spin degrees of freedom in the $s$-wave superconductor~\cite{Fidkowski11b}. Here, $\Delta_P$ and $a$ are the Cooper-pair-hopping amplitude and the theory's short distance cutoff, respectively.

As follows from Eqs.~\eqref{eq:psi-j} and \eqref{eq:psi}, the fields $\partial_x \phi_{\rho/j}$ represent the total particle number in the superconductor and semiconductors, respectively.  These field definitions lead to periodic boundary conditions, thereby simplifying the instanton calculation of the hybridization energy below (by avoiding twisted boundary conditions due to phases of the form $\exp\left\{ik_F L\right\}$).  As such, the density of the wires is fixed explicitly in the Hamiltonian by including shifts of $\partial_x\phi_j$ by $k_F$,  and of $\partial_x \phi_\rho$ by $k_F^{(\rho)}$.

In order to obtain low-energy effective description, we first run the renormalization group (RG) procedure. The superconducting pairing term is relevant and flows to strong coupling according to
\begin{equation}
\frac{dy}{dl}=\left(2-\frac{1}{2K_\rho}-\frac{1}{K}\right)y,
\end{equation}
where $y=\Delta_P a/\tilde{v}$, the length scale $l$ is defined in terms of the short distance cutoff $a_0$ as $l=\log(a/a_0)$.
We define the coherence length $\xi$ as the length scale for which $y(l)=1$, implying
\begin{equation}
\xi = a_0 \left(\frac{\tilde{v}}{\Delta_P a_0}\right)^{\left(2- \left(2K_\rho\right)^{-1}-\left(K\right)^{-1}\right)^{-1}},
\end{equation}
where the effective Fermi velocity is given by
\begin{equation}
{\tilde{v}=v_\rho/2K_\rho+v/K}.
\end{equation}

In the following, we will work in the strong-coupling limit, for which the RG is carried out until the short distance cutoff {$a\to \xi$.}  We take the mean field limit of this model to be when the velocities $v$ and $v_\rho$ are unchanged, the semiconductor is non-interacting ($K\to 1$), and the superconductor has an infinite number of channels ($K_\rho \to \infty$)~\cite{Lin98, Emery99}.  Taking this limit, we recover the mean field expressions: $\tilde{v}\to v$ and $\xi_{\text{MF}}\equiv \xi(K_\rho,K=1)|_{K_\rho \to \infty} = v/\Delta_P$.

At this scale ($a \rightarrow \xi$), one can neglect spatial fluctuations of the fields $\theta_{j/\rho}$ and take into account only uniform temporally-fluctuating modes.  Integrating out the $\phi_{j/\rho}$ fields, we have
\begin{align}\label{eq:L1-N}
\mathcal{L}^{(j)} =\frac{L}{2\pi} & \Big\{ \frac{K}{v}\left(\partial_\tau \theta_j-i \frac{v}{K} k_F\right)^2 +\frac{K_\rho}{v_\rho} \left(\partial_\tau \theta_\rho -i \frac{v_\rho}{K_\rho} k_{F}^{(\rho)}\right)^2
\\ &- \frac{\Delta_P}{\xi} \cos \left(\sqrt{2}\theta_\rho-2\theta_j\right) \Big\}\notag.
\end{align}
For the topological wire ($j$), we define average and difference fields between the nanowire and superconducting shell to be
\begin{align}
\theta_j^{+}&=  \frac{1}{2} \left(\frac{1}{\sqrt{2}}\theta_\rho+\theta_j\right)
\\ \theta_j^{-} &= \frac{1}{\sqrt{2}}\theta_\rho-\theta_j.
\end{align}
In terms of these fields, Eq.~\eqref{eq:L1-N} becomes
\begin{equation}
\begin{split}\label{eq:top-L}
\mathcal{L}^{(j)} &=\frac{L}{2\pi}
\Big\{  \frac{1}{2} \left[ \frac{K_\rho}{v_\rho}+\frac{K}{2v} \right] \left[ 4\left(\partial_\tau \theta_j^{+}\right)^2 +\left(\partial_\tau \theta_j^{-}\right)^2\right]
\\ &+ 2\left[\frac{K_\rho}{v_\rho}-\frac{K}{2v} \right]  \left(\partial_\tau \theta_j^{+}\right) \left(\partial_\tau \theta_j^{-}\right)
- \frac{ \Delta_{P}}{\xi}  \cos\left(2\theta_j^{-}\right)
\\ &-2i \left( \sqrt{2}k_{F}^{(\rho)}+k_F\right)\partial_\tau \theta_j^+- i \left( \sqrt{2}k_F^{(\rho)}-k_F\right)\partial_\tau \theta_j^-
 \Big\}.
\end{split}
\end{equation}

Integrating out the quadratic fields $\theta_j^{+}$ results in the effective action for $\theta_j^{-}$:
\begin{equation}\label{eq:Seff}
\begin{split}
S_{\text{eff}} =\frac{L}{2\pi} \int d\tau \left\{  \frac{1}{\tilde{v}} \left(\partial_\tau \theta_j^{-}+i\mu_-\right)^2  - \frac{\Delta_{P}}{\xi} \cos\left(2\theta_j^{-}\right)\right\},
\end{split}
\end{equation}
where
\begin{equation}
\mu_-\equiv \frac{v}{K} k_F-\frac{v_\rho}{\sqrt{2}K_\rho}k_F^{(\rho)} .
\end{equation}
The quantity $\mu_-$ can be understood as the Fermi energy of the semiconductor measured relative to the Fermi energy of the superconductor, we will henceforth refer to this as the relative Fermi energy.
We comment below on the role of $\mu_-$ in the dephasing of the topological qubit.  In the mean field limit, the superconductor's Fermi energy is fixed; as such $\mu_-$ is only determined by the Fermi energy of the semiconductor.
Recall that we are working in the limit that the gap in the (trivial) superconducting backbone is much larger than the gap in the topological sections of the qubit, so that fermion tunneling between the regions $(1)$ and $(2)$ is strongly suppressed.  For this reason, we have dropped an interwire coupling term,  ${\delta S^{(12)}\propto \int d\tau_1 d \tau_2 \partial_{\tau_1}\theta_1^{-}(\tau_1) \partial_{\tau_2}\theta_2^{-}(\tau_2)}$, which we do not expect to qualitatively change our results.

The two ground states of the system are $(\theta_1^-,\theta_2^-)=(0,0)$ or $(0,\pi)$.  The states
$(\pi,\pi)$ and $(\pi,0)$ are equivalent to, respectively, $(0,0)$ or $(0,\pi)$.  As discussed in Ref.~\onlinecite{Fidkowski11b}, symmetric and antisymmetric superpositions of the two ground state configurations of $(\theta_1^-,\theta_2^-)$ are associated with even and odd fermion parity in the two topological wires. Therefore, all information about the topological qubit (i.e., the MZM parity) is contained in the configuration of the $\theta_j^-$ fields.
The splitting between the two ground states can be obtained from an instanton
calculation in which $\theta_1^-$ winds by $\pi$ while $\theta_2^-$
remains unchanged, or vice versa. The two key features for the present
purposes are:
 (1) an instanton event takes the system between the two fermion parity states, as such the resulting degeneracy splitting can be associated with the MZM hybridization energy; and
 (2) there is a relative charge density buildup
associated with this instanton and, therefore, with the MZM parity state.

\subsection{MZM Hybridization Energy}

In an instanton/anti-instanton solution, $\theta_j^-(\tau)$ interpolates between ${\theta}_j^-(-\infty)=0$ and $\theta_j^-(\infty)=\pm \pi$, e.g.,
\begin{equation}\label{eq:inst-example}
{\theta}_1^-(\tau)=\pm \frac{\pi}{2} \left( 1+ \text{tanh}\left[\sqrt{\frac{\Delta_P \tilde{v}}{\xi}}(\tau-\tau_0)\right] \right).
\end{equation}
There are similar instantons in which the phase winds on the other topological segment.  We neglect multi-instanton solutions, since they have larger action and are, therefore, exponentially suppressed compared to the single instanton/anti-instanton solutions. There is a one-parameter family of such instanton/anti-instanton solutions, parameterized by the mid-point in imaginary time of the
instanton, $\tau_0$.  We must average over $\tau_0$ to include the effect of the entire family. Instantons and anti-instantons contribute with opposite phases (due to the $\mu_-$ term in the action) and have opposite charge [due to the $\pm$ sign in Eq.~(\ref{eq:inst-example})].

The instanton calculation results in the following expression for the degeneracy splitting:
\begin{equation}\label{eq:eps-hyb}
\varepsilon_{\text{hyb}} = A \cos\left(\frac{L \mu_-}{\tilde{v}}\right)\text{exp}\left\{ - \frac{L}{\xi_\text{MF}}f(K_\rho,K)\right\},
\end{equation}
where the dimensionless function in the exponent is $f(K_\rho,K)=\frac{2\sqrt{2}}{\pi} \sqrt{\frac{\xi_\text{MF}}{\xi}}$.
 Up to numerical prefactors of order one, the constant $A$ is given by the attempt frequency ${A=\sqrt{\Delta_P \tilde{v}/\xi}}$.  Equation~\eqref{eq:eps-hyb} is one of our main results of this section and, to our knowledge, the first reporting of the hybridization energy in a charge conserving formalism that captures both the oscillatory dependence and exponential suppression of the degeneracy splitting with length.  In the mean field limit (i.e. $K_{\rho}\rightarrow \infty$, $v_{\rho}={\rm const}$), ${\varepsilon_\text{hyb}^{\text{MF}}\sim \cos\left(k_F L\right)\text{exp}\left\{ - \frac{2\sqrt{2} L}{\pi \xi_\text{MF}}\right\}}$, which agrees with previous mean field calculations of the degeneracy splitting in a topological superconductor~\cite{Cheng09,Cheng10}.

\subsection{Charge Distribution}

To calculate the charge distribution associated with the MZMs, we first consider the charge densities in one of the semiconducting wires $\langle \rho_j\rangle=\frac{1}{\pi} \frac{K}{v} \langle \partial_t \theta_j\rangle$ and the neighboring region of the trivial superconductor $\langle \rho_\rho \rangle = \frac{\sqrt{2}}{\pi} \frac{K_\rho}{v_\rho} \langle \partial_t \theta_\rho\rangle$. In terms of $\theta^-$ fields, one finds
\begin{align}
\langle \rho_j\rangle &= - \frac{1}{\pi \tilde{v} } \langle \partial_t \theta_j^- \rangle + \frac{K v_\rho}{\pi}\left(\frac{ \sqrt{2} k_F^{(\rho)}+k_F}{2K_\rho v+v_\rho K}\right) \label{eq:rho-j}
\\ \langle \rho_\rho \rangle&= +\frac{1}{\pi \tilde{v}} \langle \partial_t \theta_j^- \rangle+\frac{2K_\rho v}{\pi} \left( \frac{\sqrt{2}k_F^{(\rho)}+k_F}{2K_\rho v+v_\rho K}\right). \label{eq:rho-rho}
\end{align}
 Only the first term on the right side of Eqs.~\eqref{eq:rho-j} and \eqref{eq:rho-rho} depends on the field configuration of $\theta_j^-$ and thus on the fermion parity of wire ($j$).
As one can see, the total charge expectation value of the system, $\langle \rho_j\rangle+\langle\rho_\rho\rangle$ is independent of the $\theta_j^-$ field and, thus, does not encode any topological information.  Instead, the MZM parity is encoded in a line dipole moment forming between the semiconductor and superconductor.
  Only environmental degrees of freedom that resolve the charge separation of this dipole moment couple to the MZM charge distribution.  We comment on the relevant distance scale for this dipole moment at the end of this section.

Equations~\eqref{eq:rho-j} and \eqref{eq:rho-rho} hold even if we extend the trivial superconducting region to infinity, corresponding to a grounded superconductor.  In the model presented in Ref.~\onlinecite{Sau11b}, the topological wire is an intrinsic $p$-wave superconductor with an odd number of channels.  The role played here by the semiconductor and superconductor is instead played by different channels.  As the corresponding wavefunctions will have different transverse profiles, the MZM overlap will result in some multipole charge distribution.

More explicitly, using the expression given in Eq.~\eqref{eq:inst-example} for the instanton contribution to $\theta_1^-$, we can calculate the MZM parity-dependent relative charge density $\langle\rho_-\rangle=\frac{1}{\pi\tilde{v}}\langle \partial_t \theta_1^-\rangle$. Approximating this charge as uniformly spread over the length of the topological section, we find
\begin{equation}\label{eq:QMZM}
\frac{\Delta Q_{\text{MZM}}}{e}= - \frac{L}{\sqrt{\xi_{\text{MF}}\xi}}\sin\left(\frac{L\mu_-}{\tilde{v}}\right) \text{exp}\left\{ - \frac{L}{\xi_\text{MF}} f\left(K_\rho,K\right) \right\}.
\end{equation}
Ultimately, we are interested in how the charge distribution associated with the MZMs couples to charge noise in the topological qubit's environment.  We expect electric field fluctuations to vary the parameters of the semiconductor ($k_F, \,v$) relative to those of the superconductor ($k_F^{(\rho)},\,v_\rho$), resulting in noise in the relative Fermi energy $\mu_-$.  One can verify that $\Delta Q_\text{MZM}/e= \partial_{\mu_-}\varepsilon_\text{hyb}$; combining this expression with Eq.~\eqref{eq:p-top} we have
\begin{equation}
\vec{p}_\text{top} = \Delta Q_\text{MZM}\frac{\partial \mu_-}{\partial E_z} \hat{z}.
\end{equation}
Importantly, we see that charge noise only couples to the topological qubit through the relative Fermi energy $\mu_-$ between the semiconductor and superconductor; total charge does not couple to the qubit state.
Note that in the above argument we have assumed that $\xi$ and $\xi_\text{MF}$ are parameters independent of $\mu_-$. Since the leading order $\mu_-$-dependence of $\Delta_P$ and $v$ tends to cancel in ratios $v/\Delta_P$ (see Appendix~\ref{app:exponent}), charge fluctuations predominantly couple to the prefactor rather than the exponential of the hybridization energy in Eq.~\eqref{eq:eps-hyb}.

From Eq.~\eqref{eq:rho-rho}, we see that including superconducting fluctuations was essential to observing the formation of a parity-dependent line dipole moment between the semiconductor and superconductor.  We now compare our results with the ones obtained using the BCS mean-field approximation, where superconducting fluctuations are not considered. Previous calculations using BCS mean-field theory concluded that there is a parity-dependent charge correction in the semiconductor only. Indeed, in the mean-field limit of our charge conserving formalism, our expression for $\Delta Q_{\text{MZM}}$ agrees with the BCS mean-field theory expressions in Refs.~\onlinecite{Ben-Shach15,Lin-C12,Dominguez17}. However, the latter-two papers do not take into account the screening of charge by the superconducting condensate, which exactly cancels the semiconducting contribution so that the total charge is independent of $\theta_j^-$. Thus, noise that couples to the total charge of the island does not contribute to dephasing of the topological qubit. Instead, we find that fluctuations in the electric field that couples to the line dipole moment at the superconductor-semiconductor interface contribute to dephasing. Previously, based on BCS mean-field theory calculations, $\Delta Q_\text{MZM}$ was either thought to be an absolute line charge~\cite{Lin-C12,Dominguez17}, or the relevant distance scale entering the line dipole moment was assumed to be on the order of the semiconducting wire's diameter, $w$~\cite{Ben-Shach15}.  Although our calculation does not include an estimation of the relevant length scale separating the charge in the semiconductor and superconductor, simulations of Al-InAs nanowires~\cite{Antipov18} indicate that there is an accumulation layer at the superconductor-semiconductor interface, resulting in a suppression of the dipole moment found in Ref.~\onlinecite{Ben-Shach15} by at least a factor of $r/w\sim 0.1$, where $r$ is the separation between the semiconductor and superconductor wavefunctions.  We therefore do not expect the charge $\Delta Q_\text{MZM}$ to be observable via charge sensing for wires satisfying $L/\xi>5$, as was suggested in Refs.~\onlinecite{Lin-C12} and \onlinecite{Ben-Shach15}.  The combination of the MZM charge distribution being interpreted as a dipole moment, and the concentration of the charge near the interface, suppresses the image charge effect discussed in Ref.~\onlinecite{Dominguez17} by a factor of $\left(r/w\right)^2$. As such, detecting the dielectric screening of the charge buildup in the topological wire is beyond current experimental reach.

\section{Dephasing of MZM Qubits}\label{sec:dephasing-times}

We now use the charge distribution $\Delta Q_\text{MZM}$ derived in the previous section to calculate the dephasing time of a topological qubit. We define the {\it pure dephasing time}, $T_2^*$, of the qubit to be the time scale over which off-diagonal elements of the qubit density matrix decay: $B^2(T_2^*)=1$, where $B^2(t)$ is given in Eqs.~(\ref{eq:dephasing}) and (\ref{eq:basic-dephasing}).  All qubit operations must occur on a faster time scale than the dephasing time, thus understanding the behavior of $T_2^*$ is critical for designing and building a working qubit.

Note that topological qubits are special in the sense that ideally there is no energy splitting between the two qubit states, thus which processes we call dephasing and which we call relaxation amounts to a choice of basis. We start by choosing the $z$-basis of the qubit as the parity of $i\gamma_1\gamma_2$ and neglect fermion tunneling between the two topological wires, thus reducing the problem to pure dephasing. We comment on relaxation processes at the end of this section.

The dephasing processes considered in this section are noise in the electromagnetic environment ($E$), coupling to phonons (ph), and finite temperature excitations ($\beta$).  We make the approximation that all noise sources are independent and write the dephasing exponent as a sum of the dephasing exponents from each noise source:
\begin{equation}\label{eq:total-B}
B^2(t) = B_E^2(t)+B_\text{ph}^2(t)+B_{\beta}^2(t).
\end{equation}
 We do not take into account disorder in our estimates of the different dephasing processes. Our results therefore represent the unavoidable intrinsic dephasing that is left even if growth and fabrication of the qubits is optimized. Given that topological qubits will likely be built from epitaxially grown nanowires with clean semiconductor-superconductor interfaces~\cite{Lutchyn17}, we expect that our estimates provide a good guideline for realistic dephasing times.

We begin by considering the effect of the electromagnetic environment on the qubit.  From Eqs.~(\ref{eq:HMZME}) and (\ref{eq:p-top}), we see that Eq.~(\ref{eq:dephasing}) becomes
\begin{equation}
B_{E}^2(t) =  |\vec{p}_\text{top}|^2
 \int_{1/t}^{\infty} d\omega \frac{\alpha_E}{\omega} \frac{\sin^2(\omega t/2)}{\left(\omega/2\right)^2},
\end{equation}
where we have used Eq.~(\ref{eq:1f-spec}) for the spectral function.  This expression is weakly dependent on the lower frequency cutoff, which we have approximated as $1/t$; essentially this choice of cutoff frequency amounts to only considering the noise remaining after a ``charge echo pulse''~\cite{Nakamura02}.  Solving for $B_{E}^2(T_{2,E}^*)=1$, we find
\begin{equation}\label{eq:T2-p}
T_{2,E}^* =  \left(|\vec{p}_\text{top}|\sqrt{\alpha_E \kappa} \right)^{-1},
\end{equation}
where $\kappa\equiv 1-\cos[1]+\sin[1]-\text{Ci}[1] \approx 0.96$.\footnote{Ci$[x]$ is the cosine integral function, defined by $\int_{x}^\infty dx \cos(x)/x.$}  We make the approximation that electric field can be related to the gate voltage (assumed to be applied directly at the side of the wire opposite to the superconducting shell) by $E_z w=V_g$, where $w$ is the diameter of the topological wire~\footnote{ This approximation makes the assumption that the superconductor screens the electric field, so that fluctuations in the electric field can only vary over the diameter of the semiconducting wire}.  We can then write the topological dipole moment as
\begin{equation}\label{eq:p-Q}
\vec{p}_\text{top}\sim \Delta Q_\text{MZM}  \left(\frac{\partial \mu_-}{\partial V_g}w \right)\hat{z}.
\end{equation}

Plugging Eq.~(\ref{eq:QMZM}) into Eqs.~(\ref{eq:T2-p}) and (\ref{eq:p-Q}), we see that if $\xi\approx \xi_\text{MF}$, the pure dephasing time grows with $L/\xi$ as
\begin{equation}\label{eq:T2-E}
T_{2,E}^*=c \frac{\xi}{L}\text{exp}\left\{ \frac{2\sqrt{2}}{\pi} \frac{L}{\xi}\right\},
\end{equation}
where $c=\left(w (\partial_{V_g}\mu_-)\sqrt{\alpha_E \kappa} \right)^{-1}$.
Simulation of a mean field InAs nanowire with radius $w=60$~nm, proximity-coupled to an Al superconducting shell estimates the relative Fermi energy to change with gate voltage as $\partial_{V_g} \mu_-\sim 0.1$~\cite{Antipov18}.  Making the approximation that electric field noise will be similar to the values reported in Refs.~\onlinecite{Petta04,Petersson10,Dovzhenko11,Shi13}, we set $\alpha_E=10$ (V/m)$^2$ (see Appendix~\ref{app:alpha}), resulting in $c\approx 40~$ns.    Our estimates for the dephasing time for different values of $L/\xi$ are reported in Table~\ref{table:dephasing-times}.   The dephasing times for long wires are predicted to be orders of magnitude larger than dephasing times of conventional charge qubits precisely because $\Delta Q_\text{MZM}$ is a small fraction of an electron charge.

\begin{table}
\begin{tabular}{|c|c|c|c|c|c|c|}\hline
		~$L/\xi$~ & ~$5$~ & ~$10$~ &   ~$20$~ &   ~$30^*$~ \\\hline
		~$T_{2,E}^*$~ & 600~ns & 30~$\mu$s & 100 ms & ~10 min ~ \\ \hline
		~$T_{2,\beta}^*$~ & 20~s & 20~s & 20~s & 20~s \\  \hline
		~$T_2^*$ & 200~ns & 30~$\mu$s & 100 ms & 20 s\\ \hline
\end{tabular}
\caption{
Dephasing times for the parameters of bulk InAs evaluated at different values of $L/\xi$ for different noise sources.  The first row is the pure dephasing time due solely to $1/f$ charge noise, $T_{2,E}^*$, which grows exponentially with wire length, see Eq.~(\ref{eq:T2-E}).  
The second row is the pure dephasing time due solely to thermally-excited quasiparticles in the superconductor, $T_{2,\beta}^*$, which is independent of $L/\xi$ in thermal equilibrium.  The latter only becomes relevant for long wires.  The last row is the pure dephasing time due to all three noise sources discussed in Section~\ref{sec:dephasing-times}.  We do not define a dephasing time due solely to coupling to phonons as ${B_\text{ph}^2(t)<1}$ for experimentally reasonable time scales, see Eq.~(\ref{eq:B-sq-ph}); coupling to phonons shifts the dephasing time for short wires, ${L/\xi=5}$, but has negligible effect for longer wires.  The time estimates in the table do not take into account corrections due to disorder or non-equilibrium quasiparticles in the superconductor.  The asterisk on the last column, $L/\xi\sim 30$ indicates that these corrections are likely to become important once the dephasing time estimate from intrinsic physics of the qubit (finite size effects, phonons, thermal quasiparticle excitations) has reached the order of tens of seconds.  
}
\label{table:dephasing-times}
\end{table}

In addition to $1/f$ charge noise, we can also consider dephasing from phonons coupling to the charge distribution in the MZM qubit:
\begin{equation}
B^2_\text{ph}(t) = \int_0^{\omega_D} d\omega S_\text{ph}(\omega) \frac{\sin^2(\omega t/2)}{(\omega/2)^2},
\end{equation}
where $\omega_D$ is the Debye frequency and the phonon spectral function at zero temperature can be approximated by (see Appendix~\ref{app:phonons})
\begin{equation}
S_\text{ph}(\omega) = \left( \frac{\Delta Q_\text{MZM}}{e}\right)^2 \frac{1}{(2\pi)^2 \rho v^3} \left( \frac{D^2}{v^2} \omega^3+(eh_{14})^2 M_{ii}\omega\right),
\end{equation}
where $\rho$ is the mass-density of the semiconductor, $v$ is the average of the phonons' longitudinal and transversal velocities, $D$ is the deformation potential, $h_{14}$ is the piezo-electric coupling, and $M_{ii}$ is an order one numerical factor that depends on the nanowire or quantum well geometry.
At long times, for $\xi\approx \xi_\text{MF}$, the phonon contribution to Eq.~(\ref{eq:total-B}) grows as
\begin{equation}\label{eq:B-sq-ph}
B_\text{ph}^2\left( t\gg 1 \text{s} \right) \approx \left(\frac{L}{\xi}\right)^2 \text{exp}\left\{-\frac{4\sqrt{2}}{\pi}\frac{L}{\xi}\right\} \left( a + b\log \left( t \right) \right),
\end{equation}
where, for the parameters of bulk InAs, $a\approx 300$, $b\approx 0.1$ and $t$ is measured in seconds, see Appendix~\ref{app:phonons}.  Thus, for any reasonable time scales, coupling to phonons only contributes to the MZM qubit dephasing when the constant term is of order one, i.e.,  $L/\xi\lesssim 6.5$.  For longer wires, coupling of the MZM qubit to phonons has a negligible effect on the dephasing time.

Yet another source of qubit dephasing is finite temperature excitations of quasiparticles in the superconductor.  In thermal equilibrium, finite-temperature dephasing is exponentially suppressed in $\beta \Delta$ rather than in $L/\xi$ (see Appendix~\ref{app:temperature}):
\begin{equation}\label{eq:T2-T}
T_{2,\beta}^* =\tau_0\text{exp}\left\{\beta\Delta\right\}.
\end{equation}
For the electron-phonon couplings in bulk InAs, $\tau_0\sim 50~$ns.  Using a typical value of $\beta \Delta \sim 20$,  we estimate a corresponding dephasing time of $20~$s, see Table~\ref{table:dephasing-times}. We therefore conclude that in equilibrium dephasing from thermally-excited quasiparticles can be neglected until the system is deep inside the topological regime (${L/\xi \gtrsim 20}$). 
At low enough temperatures, the superconductor may not reach thermal equilibrium and $\text{exp}\{-\beta \Delta\}$ in Eq.~\eqref{eq:T2-T} is replaced by $\sqrt{2\Delta\beta/\pi}\xi n_\text{qp}$, where $n_\text{qp}$ is the density of nonequilibrium quasiparticles. Given the small size of the superconducting shell, we expect $\xi n_\text{qp} \ll 1$ which still leads to long dephasing times. The concentration of nonequilibrium quasiparticles is highly system dependent and in most cases can be avoided by properly shielding the superconductor from extrinsic excitations; as such we do not attempt to estimate the correction to the finite temperature dephasing times from nonequilibrium effects here.

Finally, we note that throughout we assumed the limit of large charging-energy protection and thus neglected extrinsic quasiparticle poisoning as a noise source. The latter could take the qubit from its ground state subspace with total fermion parity even, to an excited state subspace with total fermion parity odd.  Extrinsic quasiparticle poisoning is exponentially suppressed in the ratio of charging energy to temperature, $\sim \text{exp}\{-\beta E_C\}$, and can be ignored provided $E_C/T\gg 1$.   Note that the charging energy decreases with qubit size ($E_C\sim L^{-1}$ for nearly-linear qubits), thus we need to use suitably designed qubits to justify ignoring this contribution to the dephasing.

In the above discussion we focused on a situation for which the qubits are susceptible to dephasing, but not to relaxation. If we include interwire fermion tunneling, MZMs $\gamma_i$ and $\gamma_j$ will in general be coupled by some hybridization energy $\varepsilon_{ij}$ and the same noise sources responsible for dephasing will cause the qubit to relax to its absolute ground state.  The time scale of this relaxation is roughly given by $T_1\sim \left(\pi \alpha_E |\varepsilon_{23}+i\varepsilon_{24}|^2/\varepsilon_{12}\right)^{-1}$, see Appendix~\ref{app:master-equation}, which is longer than the dephasing time provided $\varepsilon_{12}>\varepsilon_{23},\varepsilon_{24}$.

\section{Other effects of charge noise on the MZM qubit system}\label{sec:discussion}

Both $1/f$ charge noise and phonons couple to the qubit via a relative charge buildup between the semiconducting and superconducting wires forming in the topological sections of the qubit.  This charge is exponentially suppressed in $L/\xi$, thus in the ideal limit of infinitely separated MZMs, the qubit would be immune to such noise sources.  Essentially, finite-sized wires turn the MZM qubit into charge qubits, albeit with a much weaker coupling to the environment because $\Delta Q_\text{MZM}$ is only a small fraction of an electron charge.  As such, the dephasing times predicted in Table~\ref{table:dephasing-times} are orders of magnitude larger than typical nanosecond-scale  dephasing times for conventional charge qubits~\cite{Nakamura02,Petersson10,Paladino13}.

In addition to setting the qubit coherence times $T_1$ and $T_2^*$, one might wonder whether $1/f$ charge noise could resolve the discrepancy between the predicted oscillatory behavior of the MZM hybridization energy $\varepsilon_\text{hyb}$, see Eq.~(\ref{eq:eps-hyb}) and Refs.~\onlinecite{Prada12,DasSarma12,Lin-C12,Rainis13,Ben-Shach15}, and either the lack of oscillations~\cite{Mourik12,Deng12,Churchill13,Zhang16} or the decay of oscillations with magnetic field~\cite{Albrecht16} observed in Majorana nanowire experiments.  This discrepancy has been the subject of many studies~\cite{Dominguez17,Chiu17,Liu17a,Liu17b}, but has not yet been resolved.  If Eq.~(\ref{eq:eps-hyb}) is subject to a fluctuating electric field, the argument of the cosine can be expanded as a constant plus a fluctuating piece,
\begin{equation}
\frac{L \mu_-}{\tilde{v}} \sim \frac{L \bar{\mu}_-}{\tilde{v}}+\left(\frac{L}{\tilde{v}} \frac{\partial \mu_-}{\partial V_g}w \right)\delta E_z,
\end{equation}
where we have written the average relative Fermi energy of the topological wire as $\bar{\mu}_-$ and made the same approximation as before that $E_z w=V_g$.  The second term must be of order $\pi$ to wash out the cosine oscillations.  For {$L=1~\mu$m,} ${\tilde{v}\sim 10^5}$~m/s, {$\delta E_z\sim \sqrt{10}$~V/m}, and the parameter values used in Section~\ref{sec:dephasing-times}, the second term is too small by a factor of $10^{-4}$.  We thus do not believe that charge noise can explain the lack of oscillations in present-day experiments.

Lastly, we note that $1/f$ charge noise will also couple to the MZM qubit when it is tuned into a measurement configuration involving the auxiliary quantum dot in Fig.~\ref{fig:qubit}.  As reviewed in Section~\ref{sec:setup} and Appendix~\ref{app:QD}, an electron hopping between the quantum dot and superconducting island forming the MZM qubit will have a corresponding dipole moment $\vec{p}_\text{dot}$, which couples to $1/f$ charge noise in the same manner as does the topological dipole moment $\vec{p}_\text{top}$.  The two-level MZM island-quantum dot system dephases on a time scale
\begin{equation}\label{eq:tau-2}
\tau_2^*\sim  \left(|\vec{p}_\text{dot}|\sqrt{\alpha_E \kappa} \right)^{-1}.
\end{equation}
Furthermore, since the system-environment Hamiltonian will not be diagonal in the system's energy basis, the dot-MZM island system will relax to its ground state on a time scale set by the MZM island charging energy $E_C$ and the parity-dependent tunneling amplitude $\tilde{t}$ between the quantum dot and MZM island:
\begin{equation}\label{eq:tau-1}
\tau_1\sim  \left(\frac{4 |\tilde{t}|^2}{E_C^2} |\vec{p}_\text{dot}|^2 \frac{\pi \alpha_E}{E_C}\right)^{-1}.
\end{equation}
For $|\tilde{t}|/E_C\sim 0.1$, $E_C\sim 1$~K, and $d\sim 100~$nm, $\tau_1 \sim 5 ~\mu$s and $\tau_2^*\sim~2~$ns, see Appendix~\ref{app:QD} for details.  Conversely to the dephasing time $T_2^*$ of the MZM qubit, which we want to be as long as possible, it is beneficial for $\tau_1$ and $\tau_2^*$ to be short. The time $\tau_2^*$ quantifies how quickly the environment collapses the state during a measurement. Taking into account the measurement apparatus this time scale will be even shorter. Since the MZM parity measurement relies on the dot-MZM island system being in its ground state, $\tau_1$ effectively sets a lower bound on the measurement time if in the initialization of the measurement the charge-excited state of the system is significantly populated.

\section{Conclusions}
\label{sec:conclusions}

In this paper, we investigated intrinsic contributions to dephasing of charge-protected Majorana-based qubits built from topological superconducting nanowires, shown in Fig.~\ref{fig:qubit}.  We calculated the hybridization energy between two MZMs in a charge-conserving formalism, demonstrating that the oscillatory behavior depends on the relative Fermi energy between the semiconductor and superconductor comprising  the topological nanowire.  Furthermore, we found the charge distribution resulting from the MZM overlap is a dipole moment between the line charges in the semiconductor and superconductor; the relevant length scale entering into this dipole moment is anticipated to be much smaller than the wire radius due to an accumulation layer at the semiconductor-superconductor interface.  Thus, our findings indicate that experimental detection of the charge distribution due to the MZM overlap requires much greater sensitivity than was previously suggested~\cite{Lin-C12,Ben-Shach15,Dominguez17}. 

By estimating the electrostatic environment to be similar to that in experiments on related devices~\cite{Nakamura02,Petta04,Petersson10,Dovzhenko11,Shi13}, we calculated dephasing times due to $1/f$ charge noise coupling to the dipole moment discussed in the previous paragraph.  We reported these dephasing times in Table~\ref{table:dephasing-times} for different values of MZM separation.  By comparing dephasing from $1/f$ charge noise to dephasing from the dipole moment coupling to phonons and from thermally-excited quasiparticles in the superconductor, we expect that $1/f$ charge noise will be the dominant noise source for charge-protected MZM qubits.  We neglected extrinsic contributions to the dephasing times, such as disorder in the superconductor, which are beyond the scope of this paper.  We also find that during a measurement of the qubits in Fig.~\ref{fig:qubit}, $1/f$ charge noise couples to a dipole moment formed between the MZM island and the auxiliary quantum dot.  The coherence times associated with the combined quantum dot-MZM island system describe how quickly the environment measures the MZM parity.

Our results have important implications for future experiments on Majorana-based qubits.  In particular, in order to observe Rabi oscillations in either of the qubit designs shown in Fig.~\ref{fig:qubit}, for instance by coupling MZMs $\gamma_2$ and $\gamma_3$ for a fixed amount of time, it is necessary that the energy splitting satisfies ${\varepsilon_{23}T_2^*>1}$ so that multiple oscillations may be observed before the qubit dephases.  For $L/\xi=5,$ our estimate of $T_2^*\sim 200$~ns suggests that $\varepsilon_{23}$ must be greater than $5$~MHz.

\section*{Acknowledgments}

We are grateful to Andrey E. Antipov, David Clarke, Dmitry Pikulin, Jay Sau for stimulating discussions. C.K. acknowledges support from the NSF GRFP under Grant No. DGE $114085$.

\appendix

\section{Master equation derivation}\label{app:master-equation}

In this appendix, we derive explicit expressions for the pure dephasing time $T_2^*$ and the relaxation time $T_1$.  We begin by assuming that a system, with density matrix $\rho(t)$ has a weak interaction with the environment so that the Hamiltonian $H_{SE}=\frac{\sigma }{2}\otimes \Phi$ can be treated perturbatively.  We further assume the environment is in thermal equilibrium, described by density matrix $\rho_E$.  The interaction picture Heisenberg equation to second order in $H_{SE}$ is
\begin{equation}
\begin{split}
\dot{\rho}^I(t) = -\int_0^t dt' \tr_E \left( \left[ H_{SE}(t),\left[H_{SE}(t'),\rho^I(t')\otimes\rho_E\right]\right] \right).
\end{split}
\end{equation}
We can expand the double commutator and trace over the environmental degrees of freedom, yielding
\begin{equation}
\begin{split}
&-\dot{\rho}^I(t) =
\\ & \frac{1}{4}  \int_0^t dt'
\Big\{  \langle \Phi(t)\Phi(t')\rangle \left( \sigma(t)\sigma(t') \rho^I(t')- \sigma(t')\rho^I(t')\sigma(t) \right)
\\ &\quad \quad \quad+ \langle \Phi(t')\Phi(t) \rangle \left( \rho^I(t')\sigma(t')\sigma(t)-\sigma(t)\rho^I(t')\sigma(t') \right) \Big\}.
\end{split}
\end{equation}
We have written $\langle \Phi(t)\Phi(t')\rangle=\text{tr}_E \left\{ \Phi(t)\Phi(t')\rho_E\right\}$.
Provided the correlation time of the environment is short, we can approximate $\rho^I(t')\approx \rho^I(t)$, and extend the lower limit of integration to $-\infty$:
\begin{equation}
\begin{split}
\dot{\rho}^I(t) =& \frac{1}{4}\int_{-\infty}^t dt'
\\ \times \Big\{ &\langle \Phi(t)\Phi(t')\rangle \left(\sigma(t)\sigma(t')\rho^I(t) -\sigma(t') \rho^I(t)\sigma(t)\right)
\\ +& \langle \Phi(t')\Phi(t)\rangle \left( \rho^I(t)\sigma(t')\sigma(t)-\sigma(t)\rho^I(t)\sigma(t')\right) \Big\}.
\end{split}
\end{equation}
Finally, we change variables so that $t'\to t-t'$ and rewrite our equation in the Schroedinger picture.  Denote the energy basis of the system Hamiltonian by $\{\ket{m}\}$ such that $H_S\ket{m}=\varepsilon_m\ket{m}$.  Inserting resolutions of identity and  writing ${\Delta_{mn}\equiv \varepsilon_m-\varepsilon_n}$ we have
\begin{widetext}
\begin{equation}\label{eq:Born-Markov}
\begin{split}
\dot{\rho}_{sr}(t) + i\left( E_s-E_r\right)\rho_{sr}(t)= -\frac{1}{4} \sum_{mn} \int_0^\infty dt' \Big( &\langle \Phi(t')\Phi(0) \rangle \left[e^{-i\Delta_{mn}t'} \sigma_{sm}\sigma_{mn}\rho_{nr}(t) - e^{-i\Delta_{sm}t'}\sigma_{sm}\rho_{mn}(t)\sigma_{nr}  \right] +
\\ & \langle \Phi(0)\Phi(t')\rangle \left[ e^{-i\Delta_{mn}t'}\rho_{sm}(t)\sigma_{mn}\sigma_{nr} -e^{-i\Delta_{nr}t'}\sigma_{sm}\rho_{mn}(t)\sigma_{nr}\right] \Big).
\end{split}
\end{equation}
\end{widetext}

The master equation given in Eq.~(\ref{eq:Born-Markov}) is generally hard to solve.  We focus on the special case for which we can expand $\sigma$ in terms of Pauli matrices, ${\sigma=\sum_j a_j \sigma^j}$ with ${a_z\gg |a_x+ia_y|}$.  When considering the pure dephasing time, we restore the original upper limit of integration to $t$ rather than $+\infty$.  Then, we can approximate the equation for the off-diagonal density matrix elements as
\begin{equation}
\begin{split}
&\dot{\rho}_{01}(t)-i\Delta_{10}\rho_{01}(t) =
\\ & -\rho_{01}(t)\frac{ a_z^2}{2} \int_0^t dt' \left( \langle \Phi(t')\Phi(0)\rangle +\langle \Phi(0)\Phi(t')\rangle\right).
\end{split}
\end{equation}
We are generally interested in understanding how the magnitude of the off-diagonal elements decay,  given by
\begin{equation}
\begin{split}
&\frac{d}{dt}|\rho_{01}(t)|=\frac{d}{dt} \sqrt{\rho_{01}(t)\rho_{10}(t)}
\\&=-|\rho_{01}(t)| \frac{ a_z^2 }{2}\int_0^t dt' \left( \langle \Phi(t')\Phi(0)\rangle +\langle \Phi(0)\Phi(t')\rangle \right).
\end{split}
\end{equation}

We define the spectral function by Eq.~(\ref{eq:S-omega}), which may be equivalently written as
\begin{equation}
\langle \Phi(t)\Phi(0)\rangle +\langle \Phi(0)\Phi(t)\rangle =4 \int_0^\infty d\omega \cos(\omega t)S_{\Phi}(\omega).
\end{equation}
Then, our expression for the off-diagonal density matrix elements becomes
\begin{equation}
\begin{split}
\frac{d}{dt}|\rho_{01}(t)| &=  -|\rho_{01}(t)| 2 a_z^2 \int_0^\infty d\omega \frac{\sin(\omega t)}{\omega}S_{\Phi}(\omega).
\end{split}
\end{equation}
Integrating both sides results in Eqs.~(\ref{eq:dephasing}) and (\ref{eq:basic-dephasing}).

The pure dephasing time is defined by $B^2(T_2^*)=1$.  In the case of $1/f$ charge noise,
\begin{equation}\label{eq:T2-exp}
1 = a_z^2 \alpha_E \int_{2\pi/T_2^*}^{\infty} d\omega  \frac{\sin^2(\omega T_2^*/2)}{\omega(\omega/2)^2}=\left(T_2^*\right)^2a_z^2 \alpha_E \kappa,
\end{equation}
where $\kappa=1-\cos(1)+\sin(1)-\text{Ci}(1)\approx 0.96$.

The relaxation time is the time scale on which the diagonal density matrix element $\rho_{11}(t)$ decays.  If we assume $T_1\gg T_2^*$, then we can consider Eq.~(\ref{eq:Born-Markov}) on time scales for which the off-diagonal density matrix elements are negligible:
\begin{widetext}
\begin{equation}
\begin{split}
\dot{\rho}_{11}(t) =& - \rho_{11}(t) \frac{|a_x+ia_y|^2}{4}\int_0^\infty dt' \left( \langle \Phi(t')\Phi(0)\rangle e^{-i\Delta_{01}t'} +\langle \Phi(0)\Phi(t')\rangle e^{-i\Delta_{10}t'} \right)
\\ & + \rho_{00}(t)\frac{ |a_x+ia_y|^2}{4} \int_0^\infty dt' \left( \langle \Phi(t')\Phi(0)\rangle e^{-i\Delta_{10}t'} +\langle \Phi(0)\Phi(t')\rangle e^{-i\Delta_{01}t'}\right).
\end{split}
\end{equation}
Noting that $\rho_{00}(t)=1-\rho_{11}(t)$, we can rewrite the above as
\begin{equation}
\begin{split}
\dot{\rho}_{11}(t) &= -\rho_{11}(t)\frac{ |a_x+ia_y|^2}{4}\int_0^\infty dt'2\cos\left(\Delta_{10}t'\right)\left(\langle \Phi(t')\Phi(0)\rangle+\langle \Phi(0)\Phi(t')\rangle  \right)
\\ & \quad  +  \frac{|a_x+ia_y|^2 }{4}\int_0^\infty dt' \left( \langle \Phi(t')\Phi(0)\rangle e^{-i\Delta_{10}t'} +\langle \Phi(0)\Phi(t')\rangle e^{-i\Delta_{01}t'}\right).
\end{split}
\end{equation}
\end{widetext}
The last line just provides a constant term.  Plugging the spectral function into the first line, we find that the diagonal density matrix element decays as
\begin{equation}
\rho_{11}(t) = \rho_{11}(0)\text{exp}\left(-  \pi |a_x+ia_y|^2 S_{\Phi}(\Delta_{10})t \right).
\end{equation}
Defining the relaxation time to be the value of $t$ for which the argument of the exponent equals $-1$, we have
\begin{equation}\label{eq:T1-exp}
\left(T_1\right)^{-1} = \pi|a_x+ia_y|^2 S_{\Phi}\left(\Delta_{10}\right).
\end{equation}

\section{Tetron measurement}\label{app:QD}

In this appendix, we review how a MZM parity measurement is performed for the MZM qubits of Fig.~\ref{fig:qubit}. We then discuss the effect of $1/f$ charge noise on the measurement process.

To perform a two-MZM parity measurement, MZMs $\gamma_2$ and $\gamma_3$ are tunnel coupled to an auxiliary quantum dot, see Ref.~\onlinecite{Karzig17} for full details.  The idle qubit is described by
\begin{equation}
H_0 = H_\text{BCS}+E_C\left(\hat{N}_S-N_g\right)^2,
\end{equation}
where the first term is the BCS Hamiltonian and the second term is the charging energy Hamiltonian protecting the qubit from extrinsic quasiparticle poisoning.  The charging energy of the MZM qubit, $E_C$, is assumed to be large so that in the ground state the number of fermions in the island, counted by the operator $\hat{N}_S$, is the integer closest to the dimensionless gate voltage, $N_g$.

An effective Hamiltonian describing the auxiliary quantum dot with a single spinless fermion level is
\begin{equation}
H_\text{QD}=h\hat{n}_d +\varepsilon_C\left(\hat{n}_d-n_g\right)^2,
\end{equation}
where $h$ is an effective parameter coming from the orbital energies of the dot, $\varepsilon_C$ is the charging energy of the dot, $\hat{n}_d$ counts the electrons on the dot, and $n_g$ is the dot's dimensionless gate voltage.   The two lowest energies of the isolated quantum dot are
\begin{align}
\epsilon_0(n_g) &=\varepsilon_C n_g^2
\\ \epsilon_1(n_g) &=\varepsilon_C (1-n_g)^2 +h.
\end{align}

To perform a measurement, tunnel barriers are lowered so that the quantum dot and MZM qubit are coupled by a Hamiltonian
\begin{equation}
H_\text{tunn}= - \frac{i e^{-i\phi/2}}{2} \left( t_2 d^\dagger \gamma_2 +t_3 d^\dagger \gamma_3 \right) +\text{H.c.}
\end{equation}
The operators $e^{i\phi/2}$ and $d^\dagger$ add an electron to the MZM qubit and quantum dot, respectively.  Electrons can tunnel between the quantum dot and MZM $\gamma_j$ with amplitude $t_j$. The system Hamiltonian for the quantum dot-MZM island system is
\begin{equation}
H_S=H_0+H_\text{QD}+H_\text{tunn}.
\end{equation}
One can see that the energies of $H_S$ will depend on the MZM parity $i\gamma_2\gamma_3$, with eigenvalue $p_{23}$.  More specifically, at ${N_g=0}$, to lowest order in ${|\tilde{t}|/E_C}$ where ${\tilde{t}=-\frac{i}{2}(t_2-p_{23}t_3)}$, the ground state energy is
\begin{equation}
\varepsilon_-(n_g) = \epsilon_1(n_g) - \frac{|\tilde{t}|^2}{E_C + \epsilon_0(n_g)-\epsilon_1(n_g)}.
\end{equation}
when the decoupled system ($t_j=0$) begins in the state ${\ket{N_S=0}\ket{n_d=1}}$.  The charge distribution of the quantum dot can be understood as a first derivative of the ground state energy with respect to the gate voltage applied to the quantum dot:
\begin{equation}
q_\text{dot}(n_g) = e\left(n_g-\frac{1}{2 \varepsilon_C} \frac{\partial \varepsilon_-}{\partial n_g}\right).
\end{equation}
Working for simplicity at the quantum dot's charge degenerate point [i.e., $\epsilon_1(n_g^*)=\epsilon_0(n_g^*)$], the above is given by
\begin{equation}
q_\text{dot}(n_g^*) =e\left( 1- \frac{|\tilde{t}|^2}{E_C^2}\right).
\end{equation}
Therefore, the quantum dot's charge distribution will depend on the MZM parity, as such the two-MZM parity can be measured by charge sensing.  Importantly, we see this is a measurement of a ground state property of the combined MZM island-quantum dot system.

An electron tunneling between the auxiliary quantum dot and MZM island has a corresponding dipole moment ${\vec{p}_\text{dot}=q_\text{dot}\vec{d}}$, which couples to electric field fluctuations via
\begin{equation}
H_{SE}=\frac{1}{2}\vec{p}_\text{dot}\tau^z\otimes \delta \vec{E}.
\end{equation}
The Pauli matrix $\tau^z$ is written in the basis of an electron being on the quantum dot or the MZM island.
In order to understand coherence times of the quantum dot-MZM island system, we need to expand $\tau^z$ in the energy basis of the system Hamiltonian, $H_S$.

For simplicity, we will assume $n_g$ is tuned to the quantum dot's charge-degenerate point and that $N_g$ is tuned to the bottom of a charging parabola ($N_g=0$).   For fixed MZM parity $i\gamma_2\gamma_3=p_{23}$, the system has just two levels and it is straightforward to solve for the energies and eigenstates:
\begin{align}
\varepsilon_\pm - \epsilon_{0/1}(n_g^*) &= \frac{1}{2} \left(E_C \pm \sqrt{E_C^2+4|\tilde{t}|^2}\right)
\\ \ket{\pm} &= \frac{1}{\mathcal{N}_\pm} \left( E_C \mp \sqrt{E_C^2+4|\tilde{t}|^2},-2\tilde{t}\right)^T
\end{align}
where $\tilde{t}=-i/2(t_2-p_{23}t_3)$ and $\mathcal{N}_\pm$ is a normalization factor.  In this basis, we have
\begin{equation}
\vec{p}_\text{dot} \tau^z=\vec{a}\cdot \vec{\sigma},
\end{equation}
where $\vec{\sigma}$ is the vector of the identity and Pauli matrices and
\begin{align}
a_\mathbb{1}&= p_\text{dot} \frac{\tau^z_{++}+\tau^z_{--}}{2}
\\ a_z &= p_\text{dot} \frac{\tau^z_{++}-\tau^z_{--}}{2}
\\ |a_x+ia_y| &= p_\text{dot}|\tau^z_{+-}|.
\end{align}
We have written $\tau^z_{ab}$ to denote $\bra{a}\tau^z\ket{b}$, where $a,b=\pm$.  After some algebra, one finds
\begin{align}
a_z &= -q_\text{dot}d \frac{E_C}{\sqrt{E_C^2+4|\tilde{t}|^2}}
\\ |a_x+ia_y| &= 2q_\text{dot}d \frac{|\tilde{t}|}{\sqrt{E_C^2+4|\tilde{t}|^2}}.
\end{align}
Plugging these expressions into Eqs.~(\ref{eq:T2-exp}) and (\ref{eq:T1-exp}), we have expressions for $\tau_1$ and $\tau_2^*$, which agree with Eqs.~(\ref{eq:tau-2}) and (\ref{eq:tau-1}) to lowest order in $|\tilde{t}|/E_C$.

In order to infer the MZM parity $i\gamma_2\gamma_3$ from the charge distribution on the quantum dot, the combined MZM qubit-quantum dot system must be (predominantly) in the ground state.  If in the initialization of the measurement process the system transitions to an excited state, the measurement time must be long enough that it relaxes back to the ground state.  Therefore, $\tau_1$ is a lower bound on the measurement time if the process of tuning into and out of the measurement configuration is done diabatically.  Note that if the system remains in the ground state at all times, the measurement time could be shorter than $\tau_1$.

\section{Charge noise coupling to $v$ and $\Delta_{P}$}\label{app:exponent}

We now treat the dependency of the parameters $v$ and $\Delta_{P}$ on the mean-field level using
\begin{equation}
v=k_{F}\left(\frac{1}{m}-\frac{\alpha^{2}}{\sqrt{V_{Z}^{2}+\alpha^{2}k_{F}^{2}}}\right)\ \ \ \ \Delta_{P}=\frac{\alpha k_{F}\Delta_{0}}{\sqrt{V_{Z}^{2}+\alpha^{2}k_{F}^{2}}}\,.
\label{eq:vdelta}
\end{equation}
Here, $m$, $\alpha$, $k_F$ are, respectively, the effective mass, spin orbit coupling and Fermi momentum of the band that hosts the MZMs. For the $p$-wave gap we assume that the system is well inside the topological regime with Zeeman energy $V_Z\gg \Delta_P$. $\Delta_0$ denotes the induced $s$-wave pairing.

Since the chemical potential of the superconductor will not be affected by charge fluctuations we can set $\partial_{\mu_-}=\tilde{v}^{-1}{\partial_{k_F}}$. Using Eq.~\eqref{eq:vdelta} the derivative  $\partial_{\mu_-}\varepsilon_\text{hyb}$ now has three contributions. (1) The derivative of the attempt frequency is of the order $\partial_{\mu_-} \Delta_P$ and yields a contribution $Q^{(1)}_{\text{MZM}}$ which is bounded by $\varepsilon_\text{hyb}/\varepsilon_F$, with $\varepsilon_F$ being the Fermi energy of the band. (2) The contribution from the derivative of the cosine that was used for Eq.~\eqref{eq:QMZM} is $Q^{(2)}_\text{MZM}\sim \varepsilon_\text{hyb}/\delta$, where $\delta=v/L$ is the level spacing. (3) The derivative of the exponent contributes as $Q^{(3)}_\text{MZM}\sim (\varepsilon_\text{hyb}/\delta) \partial_{k_F}\xi^{-1}$. Since the leading order dependence of $v$ and $\Delta_P$ on $k_F$ cancels in $\xi=v/\Delta_P$, we find $\partial_{k_F}\xi^{-1}\sim \Delta_P/\varepsilon_F$. We therefore conclude that unless the system is close to the fine-tuned point $\sin(\mu_-/\delta)=0$, the relevant charge dipole can be estimated by $Q^{(2)}_\text{MZM}$ as stated in Eq.~\eqref{eq:QMZM}.

\section{Phonon dephasing}\label{app:phonons}

Coupling of the MZM qubit to phonons can be treated in a similar way to the electromagnetic noise.  For the sake of concreteness, we will focus on InAs devices.  We neglect phonons in the superconductor, which are expected to have a subleading contribution.  The phonon spectrum and electron-phonon coupling are reasonably well-understood in bulk InAs, which will allow us to place an upper bound on the
dephasing due to phonons, since the device geometry may place further restrictions on the phonon spectrum.  The qubit dephasing from phonons results from the interactions
\begin{multline}
H_\text{MZM-ph} = i\gamma_1 \gamma_2 \int\! \frac{d^3 q}{(2\pi)^3} \rhoMZM(-q)
\,\, \times\\
\left[D iq_j {u_j}(q) + e h_{14} {\sum_\lambda} {M_\lambda}(q) {\epsilon^\lambda_j}(q) {u_j}(q) \right], \label{eq:MZM-ph}
\end{multline}
where $u_j(q)$ is the Fourier transform of the displacement in the $j$th direction, $D=5.1$~eV is the conduction band deformation potential of InAs~\cite{Vurgaftman01} and
$h_{14} = 3.5\times10^6 $~V/cm is its piezoelectric coupling~\cite{Madelung04}.  Note that as InAs is electron-doped we do not need to consider the valence band deformation potential.
The form factor ${M_\lambda}(q)$ depends on the nanowire or quantum well geometry and is bounded from above by one; $\epsilon^\lambda_j$ are the polarization vectors.
This coupling is also of the form of Eq.~(\ref{eqn:H_SE}), where now $\sigma=i\gamma_1\gamma_2$ and the environment operator $\Phi$ is dependent on the charge distribution $\rhoMZM(q)$ associated with overlapping MZMs.

The noise due to phonons coupling to the MZM charge distribution is
\begin{multline}
\label{eq:phonon-noise}
a_z^2 S_\text{ph}(\omega) = \int\! \frac{d^3 q}{(2\pi)^3} |{\rho_\text{MZM}}(q)|^2 \,
\left\langle {u_i}(-q,-\omega) {u_j}(q,\omega) \big)\right\rangle\\
\times  \Big[D^2 {q^2} \delta_{ij} + (eh_{14})^2 \sum_{\lambda,\lambda'} M_{\lambda}(-q)M_{\lambda'}(q)
{\epsilon^\lambda_i}(-q)\epsilon^{\lambda'}_{j}(q)\Big].
\end{multline}
The correlation function $\left\langle{u_i}(-q,-\omega) {u_j}(q,\omega)\right\rangle$ is obtained from the fluctuation-dissipation theorem,
\begin{equation}\label{eq:fluc-diss}
\left\langle{u_i}(-q,-\omega) {u_j}(q,\omega)\right\rangle =
\chi_{ij}(q,\omega)(1-e^{-\beta\omega})^{-1},
\end{equation}
 where
\begin{multline}
\chi_{ij}(q,\omega) = \delta_{ij}\,\delta({\omega}^2 - {v_l^2}{q^2})/\rho\\
+ \, (\delta_{ij}-{q_i}{q_j}/{q^2})\,\delta({\omega}^2 - {v_t^2}{q^2})/\rho.
\end{multline}
The longitudinal and transverse phonon velocities are
${{v_l}\approx 4.7}$~km/s and ${v_t\approx 3.3}$~km/s, respectively.   The density of InAs is
$\rho\approx 5.7$~g/cm$^3$.

We approximate the charge density in the semiconducting nanowire $\rho_\text{MZM}$ as
\begin{equation}
\rho_\text{MZM}({\bf x}) = \frac{Q_\text{MZM}}{e} \frac{\delta(x)\delta(y)}{L},
 \end{equation}
with Fourier transform
\begin{equation}
\rho_\text{MZM}({\bf q})=\frac{\Delta Q_\text{MZM}}{e}\text{sinc}\left( q_x L\right).
\end{equation}
We are interested in an upper bound on the dephasing from phonons, so we approximate $\text{sinc}(q_xL)$ by 1. Thus the coupling constant $a_z$ can be identified as the dimensionless charge $Q_\mathrm{MZM}/e$.
We ignore the difference between longitudinal and transverse phonon velocities and replace $v_l$ and $v_t$ by their average, $v=4~$km/s.  Then, the spectral function of phonons coupled to the MZMs can be bounded by the expression
\begin{equation}
 S_\text{ph}(\omega)= \left[ D^2 \frac{\omega^2}{v^2} +(eh_{14})^2 M_{ii}\right] \frac{1}{\rho}  \frac{1}{(2\pi)^2} \frac{\omega}{v^3}\left(1-e^{-\beta\omega}\right)^{-1},
\label{eq:S_ph}
\end{equation}
where we have written the form factor-dependent sum in Eq.~(\ref{eq:phonon-noise}) as $M_{ij}$.  Then, dephasing from phonons is described by
\begin{equation}\label{eq:phonon-B}
\begin{split}
&B^2_{\text{ph}}(t)= \int_0^{\omega_D} d\omega \frac{\Delta Q_\text{MZM}^2}{e^2} S_\text{ph}(\omega) \frac{\sin^2(\omega t/2)}{\left(\omega/2\right)^2}
\\ =& \frac{\Delta Q_\text{MZM}^2}{(2\pi)^2 e^2 v^5 \rho} (1-e^{-\beta\omega})^{-1}
\\ &\times
 \Big(D^2\frac{ \left[2 +\left(\omega_D t\right)^2 - 2\cos\left(\omega_D t\right) - 2\omega_D t \sin(\omega_D t) \right] }{t^2}
\\ &\quad   + 2 v^2\left(e h_{14}\right)^2 M_{ii} \left[\gamma -\text{Ci}(\omega_D t) +\log(\omega_D t) \right]  \Big),
\end{split}
\end{equation}
where the Debye frequency in InAs is $\omega_D=3.3~$THz.  In the zero temperature, long time limit, $B_\text{ph}^2(t)$ grows in time as a logarithm,
\begin{equation}
\begin{split}
&B_\text{ph}^2(t\to \infty) =
\\ & \frac{\Delta Q_\text{MZM}^2}{(2\pi)^2e^2 v^5 \rho} \left( D^2 \omega_D^2 +2(e h_{14})^2 M_{ii} v^2 [\gamma+\log(\omega_D t)]\right).
\end{split}
\end{equation}
Using the upper bound $M_{ii}=3$, the above expression for the parameters of InAs is roughly
\begin{equation}
B_\text{ph}^2(t\to \infty) \approx e^{-\frac{4\sqrt{2}}{\pi}\frac{L}{\xi}} \left(\frac{L}{\xi}\right)^2 \left(300+0.1 \log[ t(1\text{ Hz})]\right).
\end{equation}
The logarithmic term only becomes important on astronomically-long time scales, thus we can judge whether phonons contribute to MZM qubit dephasing by how close the constant in time term gets to unity.  When $L/\xi>6$, $B_\text{ph}^2(t)<0.1$, and we conclude that coupling to phonons has a negligible effect on the MZM qubit dephasing. \\

\section{Dephasing due to finite temperature excitations}
\label{app:temperature}

In contrast to the discussion in Appendix~\ref{app:phonons} where phonons couple to the exponentially-small dipole moment $\vec{p}_\text{top}$, at finite temperature phonons can also lead to dephasing that is not exponentially suppressed in $L/\xi$.  Here we consider the emission of a quasiparticle from the MZMs into the continuum by absorbing a phonon from the finite-temperature bath. Such a process would takes the qubit outside of its Hilbert space and would contribute to dephasing. The corresponding timescales $T_{2,\beta}^*$ will in general be exponential in the ratio of the topological superconducting gap to the temperature, $\beta \Delta$.

Similar to Appendix~\ref{app:phonons} we first consider the effect of phonons in InAs.
The relevant part of the electron-phonon Hamiltonian that describes excitations of a MZM $\gamma$ to the continuum modes $c_k$ (with energy $\varepsilon_k>\Delta$) is

\begin{equation}
H_{\text{ex,ph}}=\int \frac{d^3q}{(2\pi)^3} \sum_k m_k(-q)(c_k^\dagger+c_k)\gamma \phi_\text{ph}(q)\,.
\end{equation}
Here $m_k(-q)=\int d^3x \psi^*_k(\textbf{x})\psi_0(\textbf{x})\exp(-i\textbf{q} \textbf{x})$ is the matrix element in terms of the (3D) wavefunction of the excited quasiparticle $\psi_k(\textbf{x})$ and MZM $\psi_0(\textbf{x})$ respectively. From Eq.~(\ref{eq:MZM-ph}), we have ${\phi_{\text{ph}}(q)=\left[D iq_j {u_j}(q) + e h_{14} {\sum_\lambda} {M_\lambda}(q) {\epsilon^\lambda_j}(q) {u_j}(q) \right]}$. To estimate an upper bound on the excitation rate, we assume ${m_k(q)=m_k\sim \sqrt{\xi/L}}$. The golden rule expression for the rate of exciting a quasiparticle $c_k$ then takes the form
\begin{equation}
\Gamma_{\gamma\to c_k}=|m_k|^2 \int dt e^{-i \varepsilon_k t}\langle \Phi_{\text{ph}}(t) \Phi_{\text{ph}}(0)\rangle\,
\end{equation}
where $\Phi_{\text{ph}}=\int d^3q \phi_{\text{ph}}(q)$. One can estimate $\Gamma_{\gamma \to c_k}$ using the phonon spectral function $S_{\text{ph}}(-\varepsilon_k\approx -\Delta)$ of Eq.~\eqref{eq:S_ph} where the appropriate coupling constant is now $m_k$. Summing over all possible excited quasiparticles (assuming a BCS-like density of states) yields
\begin{equation}
T_{2,\beta}^{*-1}=\sqrt{\frac{\pi}{2 \Delta\beta}}S_\mathrm{ph}(-\Delta)\,.
\end{equation}
Using the values of Section~\ref{app:phonons} and $\Delta=1$K, $\beta^{-1}=50$mK we find $T_{2,\beta}^*= \tau_0\exp(\beta \Delta)\approx 20\text{s}$ with $\tau_0\sim 50\text{ns}$.

In the presence of a larger-than-thermal density of  non-equilibrium quasiparticles the dominant dephasing process is due to quasiparticle relaxation into the MZMs ${T_{2,\text{neq}}^{*-1}=\sum_k n_k \Gamma_{c_k\to\gamma}}$ 
with $n_k$ denoting the occupation of the $k$th quasiparticle and $\Gamma_{c_k\to\gamma}= |m_k|^2 S_\text{ph}(\varepsilon_k\approx\Delta)$. Using the same assumptions as above we find
\begin{equation}
T_{2,\text{neq}}^{*-1}=\xi n_\text{qp} S_\text{ph}(\Delta),
\end{equation}
where $n_\text{qp}=\frac{1}{L}\sum_k n_k$ is the density of above-gap quasiparticles in the system. Since the phonon bath is in thermal equilibrium $S_\text{ph}(\Delta)/S_\text{ph}(-\Delta)=\exp(\beta \Delta)$ and we can therefore extend Eq.~\eqref{eq:T2-T} of the main text to the regime of non-equilibrium quasiparticles by identifying $\sqrt{\pi/(2\Delta\beta)}\exp(-\beta \Delta) \to  \xi n_\text{qp} $.

So far we considered only the contribution of phonons in the semiconductor assuming that most of the MZM wavefunction weight is in the semiconductor. In the case when the tunneling rate between the superconductor and semiconductor is large (i.e. strong tunneling regime), transitions due to phonons in the superconductor might become important. One can estimate the corresponding rate for Aluminum using $\tau_0^\text{(Al)}\sim 100-500$ns~\cite{Lutchyn07} and the corresponding value for $\Delta^\text{(Al)}$. Since Aluminum has weak electron-phonon coupling with $\tau_0^\text{(Al)}>\tau_0$ as well as $\Delta^\text{(Al)}>\Delta$, we expect that the excitation rate is determined by the semiconductor contribution. One can estimate an upper bound for $T_{2,\beta}^{-1}$ by assuming that most of the MZM wavefunction resides in InAs. The resulting time scale $T_{2,\beta}$ is $\sim20\text{s}$, see Table~\ref{table:dephasing-times}.

\section{Extracting $\alpha_{E}$}
\label{app:alpha}

We now explain our choice of $\alpha_E\sim 10~(\text{V/m})^2$.  Reference~\onlinecite{Petersson10} reports the spectral function describing noise for a semiconductor charge qubit as 
\begin{equation}
\label{eq:alpha-conversion}
S(\omega)=\left(\frac{E_C}{e}\right)^2\frac{ \alpha}{|\omega|},
\end{equation}
 with $\alpha=(2\times 10^{-4}~e)^2$.  In order to describe electric field fluctuations, we convert the coefficient $\alpha$ to $\alpha_E=(E_c/e)^2 \alpha/(ed)^2$, where $ed$ is the dipole moment of the double quantum dot forming the charge qubit.  Using the values $E_C=3.2$~meV and $d=200$~nm, we find $\alpha_{E}=10~(\text{V/m})^2$. 

Experiments on similar systems~\cite{Petta04,Dovzhenko11,Shi13} do not report the charging energy, but rather report $\sigma_\epsilon=\sqrt{2\int_{\omega_0}^{\omega_c}d\omega S(\omega)}$.  Assuming that the spectral function has the form of Eq.~\eqref{eq:alpha-conversion}, we calculate $\alpha_E$ for each of these papers, see Table~\ref{table:alpha}.  The bottom three rows corresponding to the more recent experiments are all of the order $10~(\text{V/m})^2$.

\begin{table}[h!]
\begin{tabular}{|c|c|c|c|}\hline
		~Ref. ~ &~ $\sigma_\epsilon$~ & ~$d$~ & ~$\alpha_E$\\ \hline
		~Ref.~\onlinecite{Petta04} ~& ~$10.2~\mu$eV ~& ~200~nm~ & ~$86~(\text{V/m})^2$~   \\ \hline
		Ref.~\onlinecite{Petersson10} & $3.9~\mu$eV & 200~nm &~$13~(\text{V/m})^2$ ~\\ \hline
		Ref.~\onlinecite{Dovzhenko11} &$7.3~\mu$eV & 250~nm & $28~(\text{V/m})^2$ \\ \hline
		Ref.~\onlinecite{Shi13} & $5~\mu$eV & 200~nm  & $21~(\text{V/m})^2$  \\ \hline
\end{tabular}
\caption{
Values of $\alpha_E$ for Refs.~\onlinecite{Petta04,Petersson10,Dovzhenko11,Shi13} using frequency cutoffs $\omega_0=2\pi/(100~\text{ms})$, $\omega_c=40$~MHz reported in Ref.~\onlinecite{Petersson10}.
}
\label{table:alpha}
\end{table}

\bibliography{topo-phases}

\end{document}